\documentclass[12pt]{article}
\usepackage{amsmath}
\usepackage{amssymb}
\usepackage{jheppub}
\usepackage{amsmath, amsthm, amsfonts, amssymb, bbm}
\usepackage[mathscr]{eucal}
\usepackage{graphics}
\usepackage{xypic}
\usepackage[all]{xy}\xyoption{rotate}
\usepackage[english]{babel}
\usepackage[font=small,format=plain,labelfont=bf,up]{caption}
\usepackage[bbgreekl]{mathbbol}
\usepackage{tikz}

\title{Logarithmic W-algebras and Argyres-Douglas theories at higher rank }

\author{Thomas Creutzig}

\affiliation{Department of Mathematical and Statistical Sciences, University of Alberta,
Edmonton, Alberta  T6G 2G1, Canada and \newline
Research Institute for Mathematical Sciences, Kyoto University, Kyoto Japan 606-8502 \newline 
email: creutzig@ualberta.ca}

\date{}

\begin{document}

\abstract{
Families of vertex algebras associated to nilpotent elements of simply-laced Lie algebras are constructed. These algebras are close cousins of logarithmic W-algebras of Feigin and Tipunin and they are also obtained as modifications of semiclassical limits of vertex algebras appearing in the context of $S$-duality for four-dimensional gauge theories.
In the case of type $A$ and principal nilpotent element the character agrees precisely with the Schur-Index formula for corresponding Argyres-Douglas theories with irregular singularities. For other nilpotent elements they are identified with Schur-indices of type IV Argyres-Douglas theories. 
Further, there is a conformal embedding pattern of these \voas{} that nicely matches the RG-flow of Argyres-Douglas theories as discussed by Buican and Nishinaka.
}

\theoremstyle{plain}
\newtheorem*{introthm}{Theorem}
\newtheorem{obs}{Observation}
\newtheorem{thm}{Theorem}[section]
\newtheorem{prop}[thm]{Proposition}
\newtheorem{lem}[thm]{Lemma}
\newtheorem{cor}[thm]{Corollary}
\newtheorem{conj}[thm]{Conjecture}

\theoremstyle{definition}
\newtheorem{defi}[thm]{Definition}
\newtheorem{rem}[thm]{Remark}

\newcommand {\AL}{\mathbb L}
\newcommand {\CC}{\mathbb{C}}
\newcommand {\cC}{\mathcal{C}}
\newcommand {\ZZ}{\mathbb{Z}}
\newcommand {\tr}{\text{tr}}
\newcommand {\ch}{\text{ch}}
\newcommand {\sch}{\text{sch}}
\newcommand {\sltwo}{\mathfrak{sl}_2}
\newcommand {\cW}{\mathcal{W}}
\newcommand {\cM}{\mathcal{M}}
\newcommand {\cX}{\mathcal{X}}
\newcommand {\cB}{\mathcal{B}}
\newcommand {\cF}{\mathcal{F}}
\newcommand {\g}{\mathfrak{g}}
\newcommand {\cS}{\mathcal{S}}
\newcommand {\cO}{\mathcal{O}_p}
\newcommand {\voa}{vertex operator algebra}
\newcommand {\voas}{vertex operator algebras}
\newcommand{\Sing}{M(p)}
\newcommand{\Trip}{W(p)}

\newcommand{\hopflink}{{\text{\textmarried}}}

\renewcommand{\baselinestretch}{1.2}

\maketitle

\section{Introduction}

The main point of this work is to construct \voas{} whose characters coincide with Schur indices of corresponding Argyres-Douglas theories as computed by Buican and Nishinaka \cite{BN}. Moreover the construction makes the predicted automorphism group $\mathbb Z/2\mathbb Z \times S_N$ manifest and also gives a nice explanation of RG-flow via conformal embeddings. At the same time these \voas{} are also interesting from the point of view of logarithmic conformal field theories. 

The slogan initiated in \cite{Beem:2013sza} is that interesting quantities of higher dimensional gauge theories are sometimes described by \voas. There are various manifestations of this picture and the one that I am familiar with is $S$-duality for four-dimensional $\mathcal N=4$ supersymmetric GL-twisted gauge theories. There the two-dimensional intersection of three-dimensional topological boundary conditions is described by certain families of \voas, parameterized by the gauge coupling. Moreover categories of line defects ending on the boundary conditions are categories of modules of the corner \voa, see \cite{Creutzig:2017uxh, Gaiotto:2017euk}. There is no apparent relation known to me between these supersymmetric gauge theories and Argyres-Douglas theories\footnote{Enhancement of supersymmetry can be explained via RG-flow as in \cite{Buican:2018ddk}}.
 Nonetheless, one construction of this work is a modification of a semi-classical limit of \voas{} arising naturally in this $S$-duality picture. 

Argyres-Douglas theories are four-dimensional $\mathcal N=2$ supersymmetric field theories \cite{Argyres:1995jj,Argyres:1995xn,Eguchi:1996vu} and they are very interesting from the \voa{} perspective as their Schur indices are believed to coincide with characters of certain \voas{} \cite{Xie:2016evu, Buican:2015ina, Cordova:2017mhb, Cordova:2017ohl, C}. Moreover indices of line and surface defects are related to characters of modules of the \voa{} \cite{Cordova:2017mhb, Cordova:2017ohl}. Interestingly also modularity and Verlinde-like formulae have nice interpretations in the gauge theory \cite{Fredrickson:2017yka, Neitzke:2017cxz, Kozcaz:2018usv}. For some further recent progress see \cite{Buican:2017fiq, Ito:2017ypt, Agarwal:2017roi,Fluder:2017oxm, Giacomelli:2017ckh,Choi:2017nur, Buican:2016arp, Buican:2017rya}.

Personally, I find it very interesting that the \voas{} appearing in this gauge theory context are often of logarithmic type, that is they allow for indecomposable but reducible representations. There are various difficult questions in the context of logarithmic \voas, see \cite{Creutzig:2013hma} for an introduction, and it is good that they allow for gauge theory interpretations. So far the best studied logarithmic \voas{} are the triplet algebras \cite{Feiginetal,Am1,AM2,Tsuchiya:2012ru}, the fractional level WZW theories of $\mathfrak{sl}_2$ \cite{Creutzig:2012sd,Creutzig:2013yca,Ridout:2008nh}, the logarithmic $\cB(p)$-algebras  \cite{CRW} and some progress is currently made on higher rank cases \cite{Kawasetsu:2018lur}. All these examples have appearances in higher dimensional super conformal field theories.
Especially, I would like to point out the work on the representation theory of the logarithmic $\cB(p)$-algebras \cite{ACKR}. These correspond to Argyres-Douglas theories of type $(A_1, A_{2p-3})$ and it might serve as a nice example for further investigation of the relation between representation theoretic data of the vertex algebra and properties of the gauge theory. Note that these representation theories usually have uncountable many inequivalent simple objects and in order to go to a conjectural modular tensor category one has to look at the semisimplification, that is the category where negligible morphisms are identified with the zero morphisms. For details and conjectures in the \voa{}\ context on this please see work with Terry Gannon  \cite{Creutzig:2016fms}.
Let me now state a few examples of the connection between representation theory data and super conformal field theory data. 
\begin{enumerate}
\item The paper \cite{Fredrickson:2017yka} observes for some eamples of type $A_1$ Argyres-Douglas theories that a limit of wild Hitchin characters can be expressed in terms of the modular data of the semisimplification of the module category of the corresponding \voa. 
\item In \cite{Neitzke:2017cxz} the operator algebra of $\frac{1}{2}$-BPS line defects was studied and a main claim is that the homomorphism from this operator algebra to the algebra of functions on $U(1)_r$-invariant vacua of the theory compactified on $S^1$ factors through the Verlinde algebra of the semisimplification of the category of modules of the chiral algebra. 
\item
Prototypical examples of negligible modules, i.e. modules on which the identity is a negligible morphism, are the relaxed-highest weight modules of affine vertex algebras, see e.g. \cite{Kawasetsu:2018lur}. If we restrict to $\sltwo$ at admissible level or to the $\beta\gamma$ \voa, then characters of these modules are formal distributions times certain modular functions \cite{Creutzig:2012sd,Creutzig:2013yca,Ridout:2008nh}. It turns out that indices of surface defects can be identified with characters of relaxed-higest weight modules in certain examples \cite{Cordova:2017mhb}\footnote{The authors of \cite{Cordova:2017mhb} denote relaxed-highest weight modules bilateral modules. }.
\end{enumerate}

\subsection*{Results}

Buican and Nishinaka give a nice compact  formula for Schur-indices of $(A_{N-1}, A_{N(n-1)-1})$ Argyres-Douglas theories \cite{BN}. The first statement is a rewriting of this formula that makes comparison with \voa{} characters evident. 
The resulting expression is the outcome of section \ref{sec:Schur} and it reads
\begin{equation}
\mathcal I(N, n)  = \frac{q^{\frac{\ell}{12}+\rho^2}}{\eta(q)^{2\ell}}\sum_{\lambda \in P^+} \sum\limits_{w\in W}\epsilon(w)q^{- (\rho, w(\lambda+\rho))}q^{\frac{n}{2}(\lambda, \lambda+2\rho)} \sum_{\mu \in Q+\lambda} m_{\lambda}(\mu) q^{-\frac{n}{2}\mu^2} x^\mu.
\end{equation}
Here $\g=\mathfrak{su}(N)$ and $\ell$ is the rank of $\g$, $P^+$ its set of dominant weights, $W$ the Weyl group of $\g$, $\epsilon(w)$ the parity of the Weyl reflection $w$, $\rho$ is the Weyl vector, $Q$ the root lattice and $m_\lambda(\mu)$ is the dimension of the weight space of weight $\mu$ in the irreducible highest-weight representation $\rho_\lambda$ of highest-weight $\lambda$ of $\g$. 

In section \ref{sec:Bp} a family of \voas{} that are denoted $B(n)_Q$-algebras are constructed. Here $Q$ is the root lattice of a simply-laced Lie algebra and $n>1$ is a positive integer larger than one. The construction is a generalization of joint work with Ridout and Wood \cite{CRW} and it is a modification of certain logarithmic W-algebras of Feigin and Tipunin \cite{Feigin:2010xv}. The main result is then that for $Q=A_{N-1}$
\begin{equation}
\ch[\cB(n)_{A_{N-1}}](q,x) = q^{-\frac{c}{24}} \mathcal I(N, n)(q, x)
\end{equation}
with the central charge of the $\cB(n)_Q$-algebra 
\[
c= 2\ell - \frac{\rho^2}{2n}(n-1)^2,
\]
and we check that this precisely matches the proposed central charge (see \cite{Xie:2012hs}) of the expected chiral algebra for the Argyres Douglas theory.

In section \ref{sec:morevoas} a more general but conjectural construction is given. The idea is to first take a semi-classical limit of \voas{} conjectured by Davide Gaiotto and myself in \cite{Creutzig:2017uxh}. The resulting \voas{} are  certain objects in a completion of a category of $G\otimes W(\g, f)$-modules with $W(\g, f)$ the quantum Hamiltonian reduction associated to $f$ and $\g$ at a certain level and $G$ the compact Lie group of the simply-laced Lie algebra $\g$.  The idea is now that one can replace Rep(G) by a braided equivalent tensor category of modules for the Heisenberg \voa{} of rank the rank of $\g$. Essentially we forget the $G$ action to view $G$-modules as $T$-graded vector spaces with $T$ the maximal torus of $G$ and $T$-graded vector spaces are equivalent to categories of Fock modules of the Heisenberg \voa, see e.g. section 2 of \cite{Creutzig:2017khq}.
We then restrict to $\g=\mathfrak{sl}_N$. For $f$ prinipal nilpotent we then recover the \voa{} of our previous construction and for general $f$ we get more new \voas. Their characters can be identified with type IV Argyres-Douglas theories which are labelled by Young tableaux and as explained in section 4 of \cite{Beem:2014rza} super conformal indices change according to the Euler-Poincar\'e principle for the corresponding quantum Hamiltonian reduction. 

One nice outcome of this construction is that all \voas{} have as manifest symmetry the Weyl group $W$ times the automorphism group of the corresponding W-algebra. For example in the case of $\g=\mathfrak{sl}_N$ and $f$ prinipal nilpotent this is $\mathbb Z/2\mathbb Z \times S_N$ as predicted by Buican and Nishinaka \cite{BN}.

Another nice outcome is that our explixit construction allows to study conformal embeddings of resulting \voas, see section \ref{sec;emb}. The idea is rather simple and is taken from \cite{Genra}. The new \voas{} are described as intersections of kernels of screening charge on a free field algebra. Removing one such screening, that is only considering the intersection of all screening except for one gives a bigger \voa. This results in the embedding
\[
\cB(n)_{Q_N} \subset   \cB(n)_{Q_\nu} \otimes \cB(n)_{Q_{N-\nu}} \otimes W_{D^0_{\Phi'}(n)},
\]
with the last factor some extension of a rank two Heisenberg \voa{} by Fock modules in a rank one isotropic lattice. This matches nicely the discussion of RG-flow in section 4.2 of \cite{BN} and so here we see that RG-flow on the level of \voas{} is described by conformal embeddings. 

\subsection*{Questions}

The present work gives a \voa{} realization or interpretation of the findings of Buican and Nishinaka \cite{BN}. There are a few natural questions that are interesting from physics and mathematics perspective. 
\begin{enumerate}
\item Is there a domain for the Jacobi variables so that the characters of the \voas{} found here converge in this domain and can be meromorphically continued to meromorphic Jacobi forms? The answer is yes for the case of $\g=\sltwo$, see \cite{C}. In this case characters also have a nice product form and so a second question would be if the more general case has a nice product form as well. 
\item Setting the Jacobi variable to one, one expects modular forms as characters. This property holds for characters of ordinary modules of quasi-lisse \voas{} \cite{Arakawa:2016hkg}. Quasi-lisse means that the associated variety (the maximal spectrum of the $C_2$-quotient) is symplectic with only finitely many symplectic leaves. Presumably the new \voas{} of this work are quasi-lisse. Characters and thus Schur indices would then be solutions to certain modular differential equations, see \cite{Beem:2017ooy, Arakawa:2017aon} for discussions of relations to Higgs branches.
\item The \voas{} here have been constructed as modifications of semi-classical limits of \voas{} appearing in the context of S-duality for four-dimensional supersymmetric gauge theories \cite{Creutzig:2017uxh} and one might wonder if the construction has a physics explanation, i.e. is there a relation between \cite{Creutzig:2017uxh} and Argyres-Douglas theories?
\item From the \voa{} perspective generalizations to non simply-laced and even superalgebras is possible, though probably difficult. For example it should be possible to construct a \voa{} associated to $F_4$ whose character coincides with \cite[eq. (6.7)]{BN}.
\item Level-rank type dualities for the \voas{} appearing here seem to be rather rich and deserve further study. The expectation is that Heisenberg cosets of $\cB(n)_{A_{N-1}}$ are isomorphic to certain $W$-algebras and the two $A_1$ examples corresponding to $\cB(4)_{A_{1}}$ and $\cB(5)_{A_{1}}$ of this are given in Theorem 10.5 and 6 point (2) of \cite{Linshaw:2017tvv}. 
\item Schur indices of Argyres-Douglas theories of type $(A_{n-1}, A_{m-1})$ with gcd$(n, m)\notin \{ 1, n, m\}$ are unknown. The construction of this paper can however (at least conjectural) be generalized. I plan to report on this generalization as well as its relation to level-rank duality in the future and hope that it will shed light on the understanding of chiral algebras of these unexplored Argyres-Douglas theories. 
\end{enumerate}

\noindent{\bf Acknowledgements} Most of all I appreciate very much the many instructive discussions with Takahiro Nishinaka.
I am also grateful to Matthew Buican for his interest and his comments to this work and I would like to thank Davide Gaiotto and Boris Feigin for many discussions on related issues. 
 I am supported by NSERC Discovery grant number RES0020460.

\section{Schur-indices of type $(A_{N-1}, A_{N(n-1)-1})$}\label{sec:Schur}

In \cite{BN}, Buican and Nishinaka found a nice compact formula for Schur-indices of $(A_{N-1}, A_{N(n-1)-1})$ Argyres-Douglas theories. I will rewrite them in a form that makes the comparison to \voa{} characters obvious.  
They start with a proposal for simple wave functions for certain irregular punctures in SU(N) $q$-deformed Yang-Mills theory,
\[
\tilde f_R^{(n)}(q, x) = \prod_{k=1}^\infty (1-q^k)^{1-N} q^{nC_2(R)} \tr_R\left(q^{-\frac{n}{2}F^{i,j}h_ih_j}x \right).
\]
Here $F_{i, j}$ denotes the Gram matrix of the weight lattice $P$ of $\mathfrak{su}(N)$. $N-1$ is the rank $\ell$ of it. $R$ denotes an irreducible highest-weight representation of $\mathfrak{su}(N)$, so we prefer to denote it by the corresponding highest-weight $\lambda$. $C_2(R)$ is the eigenvalue of the quadratic Casimir, which is $(\lambda, \lambda+2\rho)/2$ with $\rho$ the Weyl vector, i.e. one half the sum of all positive roots. The element $x$ is in the Cartan subalgebra. Let $m_{\lambda}(\mu)$ be the multiplicity of $\mu$ in the representation $R$ of highest-weight $\lambda$, then 
\[
\tr_R\left(q^{-\frac{n}{2}F^{i,j}h_ih_j}x \right) = \sum_{\mu \in Q+\lambda} m_{\lambda}(\mu) q^{-\frac{n}{2}\mu^2} x^\mu, \qquad x^\mu := \mu(x).
\]
The root lattice is denoted by $Q$.
So that we have rewritten
\[
\tilde f_R^{(n)}(q, x) = \frac{q^{\frac{\ell}{24}}}{\eta(q)^\ell}  q^{\frac{n}{2}(\lambda, \lambda+2\rho)} \sum_{\mu \in Q+\lambda} m_{\lambda}(\mu) q^{-\frac{n}{2}\mu^2} x^\mu
\]
for $R$ the highest-weight representation of highest-weight $\lambda$. 

The second ingrededient is the coefficient
\[
C_R(q) = \frac{\prod\limits_{k=1}^{N-1}(1-q^k)^{N-k}}{(q;q)_\infty^{N-1}} \chi_R\left(q^{-\frac{N-1}{2}}, q^{-\frac{N-3}{2}}, \dots, q^{\frac{N-1}{2}}\right).
\]
The last term needs explanation. The character is defined to be
\[
\chi_\lambda(x) = \tr_\lambda(x) = \sum_{\mu \in Q+\lambda} m_{\lambda}(\mu)  x^\mu
\]
and here \cite{BN} chose a standard realization of the simple roots of $A_{N-1}$. Namely consider $\epsilon_1, \dots, \epsilon_N$ with product $\epsilon_i\epsilon_j =\delta_{i, j}$ and then the simple positive roots are $\alpha_i=\epsilon_i-\epsilon_{i+1}$. One identifies weights with the Cartan subalgebra via the pairing so that one can write $x= x_1^{\epsilon_1} \dots x_N^{\epsilon_N}$, with $x_i\in \mathbb C$. For $x$ to be in the Cartan subalgebra
one actually has to inforce that $x_1 \cdot \dots \cdot x_N=1$ so that $\epsilon_1 +\dots +\epsilon_N$ acts trivially. 
 With this notation one has
\[
\chi_\lambda(x) =  \chi_R\left(x_1, \dots, x_N \right).
\]
Now, $x_j=q^{j-\frac{N+1}{2}}= e^{2\pi i \tau \left(j-\frac{N+1}{2}\right)}$, so that
\begin{equation}
\begin{split}
x &= x_1^{\epsilon_1} \dots x_N^{\epsilon_N} = \prod_{j=1}^N e^{2\pi i \tau \left(j-\frac{N+1}{2}\right)\epsilon_j} = e^{-2\pi i \tau \rho} = q^{-\rho}
\end{split}
\end{equation}
with the Weyl vector 
$$\rho = \sum_{\alpha\in \Delta_+} \alpha =- \sum_{j=1}^N \left(j-\frac{N+1}{2}\right)\epsilon_j.$$
Recall that $\Delta_+$ denotes the set of positive roots which are precisely the $\epsilon_{i}-\epsilon_j$ for $i<j$. The Weyl denominator is
\[
\delta(x) = x^\rho \prod_{\alpha\in \Delta_+}(1-x^{-\alpha})= x^\rho \prod_{\substack{i, j =1\\ i<j}}^N (1-x_ix_j^{-1})
\]
so that
\[
\delta(q^{-\rho}) = q^{-\rho^2}\prod_{\substack{i, j =1\\ i<j}}^N (1-q^{-(j-i)})= \prod\limits_{k=1}^{N-1}(1-q^k)^{N-k}.
\]
In summary $C_R(q)$ has been rewritten as
\begin{equation}
\begin{split}
C_R(q) &= \frac{\prod\limits_{k=1}^{N-1}(1-q^k)^{N-k}}{(q;q)_\infty^{N-1}} \chi_R\left(q^{-\frac{N-1}{2}}, q^{-\frac{N-3}{2}}, \dots, q^{\frac{N-1}{2}}\right) \\
&= \frac{q^{\frac{\ell}{24}}}{\eta(q)^\ell} q^{\rho^2}\delta(q^{- \rho})\chi_\lambda(q^{-\rho}),
\end{split}
\end{equation}
where $R$ is the highest-weight representation of highest-weight $\lambda$ and $\ell=N-1$ is the rank. 
Weyl's character formula says that
\[
\chi_\lambda(e^x) = \frac{\sum\limits_{w\in W}\epsilon(w)e^{(x, w(\lambda+\rho))}  }{\delta(e^x)}
\]
with $W$ the Weyl group and $\epsilon(w)$ the sign of the Weyl reflection $w$, i.e. minus for an odd reflection and plus for an even one. 
Thus putting everything together one gets the desired rewriting of the Schur index. The Schur index is
\[
\mathcal I(N, n):= \mathcal I_{A_{N-1, N(n-1)-1}} =   \sum_R C_R(q) \tilde f_R^{(n)}(q, x),
\]
here the sum is over all inequivalent finite dimensional simple modules of SU(N). These are parameterized by the positive Weyl chamber $P^+$. 
Thus 
\begin{equation}
\mathcal I(N, n)  = \frac{q^{\frac{\ell}{12}+\rho^2}}{\eta(q)^{2\ell}}\sum_{\lambda \in P^+} \sum\limits_{w\in W}\epsilon(w)q^{- (\rho, w(\lambda+\rho))}q^{\frac{n}{2}(\lambda, \lambda+2\rho)} \sum_{\mu \in Q+\lambda} m_{\lambda}(\mu) q^{-\frac{n}{2}\mu^2} x^\mu.
\end{equation}
The main objective is to construct \voas{} whose graded characters coincide with these Schur indices. Before doing so one can obtain a series of more general Schur indices as discussed in the introduction of \cite{BN}. These correspond to type IV Argyres-Douglas theories, see \cite{Xie:2012hs, Xie:2013jc}. They are labelled by Young tableaux which can be interpreted as partitions of the integer $N$ with $N$ the number of boxes of the Young tableau.
Another interpretation is as decompositions of the standard representation of SU(N) into irreducible representations of SU(2) and thus determining an embedding of SU(2) in SU(N). This latter is exactly what determines the different quantum Hamiltonian reductions of the affine \voa{} of $\mathfrak{sl}_n$ 
to $W$-algebras. The example discussed above corresponds to the principal embedding of SU(2) in SU(N) which corresponds to the patition $N=N$ or the Young tableau consisting just out of one column of height $N$. 
The simplest example is the partition $N= 1+  1 + \dots + 1$ corresponding to SU(N) decomposing into $N^2-1$ copies of the trivial representation of SU(2). In this case the $W$-algebra is just the affine \voa, i.e. no reduction is performed. The Schur-index is then given by
\begin{equation}\label{eq:01}
 \mathcal I_{A_{N-1, N(n-1)-1}, \Box\Box\cdots\Box} =   \sum_R C^{\Box\Box\cdots\Box}_R(q, z) \tilde f_R^{(n)}(q, x)
\end{equation}
with $C^{\Box\Box\cdots\Box}_R(q, z)$ defined as 
\begin{equation}
C^{\Box\Box\cdots\Box}_R(q, z) = P.E.\left(\frac{q}{1-q} \chi_{\text{adj}}(z)\right) \chi_R(z)
\end{equation}
and $\chi_R$ the character of the representation $R$ and 
\[
P.E.(f(q, z)) = \exp\left(\sum_{n=1}^\infty f(q^n, z^n)\right).
\]
So that $C^{\Box\Box\cdots\Box}_R(q, z)$ is nothing but the $\chi_R(z)$ times the Weyl denominator of $\widehat{\mathfrak{sl}}_n$,
\begin{align} \nonumber
%\begin{split} 
C^{\Box\Box\cdots\Box}_R(q, z)&= P.E.\left(\frac{q}{1-q} \chi_{\text{adj}}(z)\right) \chi_R(z) \\ \label{eq:02}
&= \exp\left(\sum_{n, m=1}^\infty q^{nm} \left( \ell + \sum_{\alpha\in \Delta_+}\left(z^{n\alpha}+z^{-n\alpha}\right) \right) \right)\chi_R(z) \\ \nonumber
&=\exp\left(-\sum_{m=1}^\infty \ell \ln(1-q^{m})  + \sum_{\alpha\in \Delta_+}\ln\left(1-z^{\alpha}\right)+\ln\left(1-z^{-\alpha}\right) \right) \chi_R(z) \\  \nonumber
&=\prod_{m=1}^\infty \frac{\chi_R(z)}{(1-q^m)^\ell  \prod\limits_{\alpha\in \Delta_+}(1-z^\alpha q^m)(1-z^{-\alpha}q^m)}.
%\end{split}
\end{align}
Here $\ell=N-1$ is the rank of $\g=\mathfrak{sl}_N$ and $\Delta_+$ is the set of simple roots. 
Recall that the character of the irreducible highest-weight representation $\AL_k(\lambda)$ of highest-weight $\lambda$ for generic $k$ is given by 
\begin{equation}\label{eq:03}
\ch[\AL_k(\lambda)] = q^{\frac{(\lambda, \lambda+2\rho)}{2(k+h^\vee)}-\frac{c}{24}} \prod_{m=1}^\infty \frac{\chi_\lambda(z)}{(1-q^m)^\ell  \prod\limits_{\alpha\in \Delta_+}(1-z^\alpha q^m)(1-z^{-\alpha}q^m)}
\end{equation}
so that we recognize $C^{\Box\Box\cdots\Box}_R(q, z)$ to be this character up to the prefactor $q^{\frac{(\lambda, \lambda+2\rho)}{2(k+h^\vee)}-\frac{c}{24}}$. Similarly, if we pass to a non-trivial Young tableau, then as explained in section 4.3 of \cite{Beem:2014rza}, see also \cite{Gadde:2011ik}, one has to replace this character by the Euler-Poincar\'e character of the corresponding quantum Hamiltonian reduction. For example $C_R(q, z)$ 
is obtained from $C^{\Box\Box\cdots\Box}_R(q, z)$ by multiplying it with the supercharacter of a fermionic ghost \voa{} and then specializing the Jacobi variable to $z=q^{-\rho}$.

In the following sections these procedures will be realized on the level of \voa{} constructions. 

\section{The $\cB(n)_Q$-algebra }\label{sec:Bp}

In \cite{CRW} a series of $W$-algebras named $\cB_p$-algebras were constructed. The construction is reviewed in \cite{C} and it goes as follows. The triplet \voa{} is defined as the kernel of a screening charge inside the lattice \voa{} of the rescaled root lattice $\sqrt{p}A_1$. Its $U(1)$-orbifold is called the singlet algebra \cite{Adamovic:2007qs,Creutzig:2013zza,Creutzig:2016htk}.
The $\cB_p$-algebra is a simple current extension of an U(1)-orbifold of the triplet algebra times a Heisenberg \voa, that is a free boson. Namely, consider the isotropic diagonal sublattice inside $\sqrt{p}A_1' \oplus \sqrt{-p}A_1'$ and the kernel of screening on the lattice \voa. This is the $\cB_p$-algebra and it corresponds to the Schur-index of $(A_1, A_{2(p-1)-1})$-Argyres-Douglas theories. 

There are similar higher rank analogues to the triplet and singlet algebras constructed by Feigin and Tipunin \cite{Feigin:2010xv}. These have been investigated in \cite{Creutzig:2016uqu} and I follow that notation. Let $\mathfrak g$ be a simply-laced Lie algebra and let $Q$ be its root lattice and $P$ the weight lattice. Moreover let $P_+$ be the positive Weyl chamber. 
The higher rank analogue of the triplet \voa{} is denoted by $\cW(p)_Q$ with $p\in \mathbb Z_{>1}$ and it is defined as the joint intersection of certain screening operators $e_0^{-\alpha_i/\sqrt{p}}$ associated to the simple roots $\alpha_1, \dots, \alpha_\ell$ with $\ell$ the rank of $\mathfrak g$:
\[
\cW(p)_Q := \bigcap_{j=1}^\ell \text{ker}_{V_{\sqrt{p}Q}} e_0^{-\alpha_j/\sqrt{p}}.
\]
Here $V_{\sqrt{p}Q}$ denotes the lattice \voa{} associated to the rescaled root lattice $\sqrt{p}Q$.
To match notation with the previous section I will now replace $p$ by $n$. 
The $\cB(n)_Q$-algebra will be defined as an intersection of the same screenings on a different \voa, namely on the tensor product of a certain isotropic lattice \voa{} and a Heisenberg \voa. The lattice in question is isotropic, that is every vector has norm zero. For this let $$D^\pm =\sqrt{\pm n}P = \mathbb Z \beta_1^\pm \oplus \dots \oplus \mathbb Z \beta_\ell^\pm$$ with $\beta^\pm_i \beta^\pm_j = \pm n G_{i, j}$ and $G$ the Gram matrix of $P$. We then define the isotropic lattice 
\begin{equation}\label{eq:isotropiclattice}
D^0_P(n) :=  \mathbb Z \left(\beta_1^++\beta_1^-\right) \oplus \dots \oplus \mathbb Z \left(\beta_\ell^++\beta_\ell^-\right).
\end{equation}
Let $d= \text{rank}(\g)$ and $\mathcal H(2d)$ be the Heisenberg \voa{} of rank $2d$ and let $\mathcal F_\lambda$ be the Fock module of weight $\lambda$, then the
 corresponding \voa{} is
\[
W_{D^0_P(n)} := \bigoplus_{\lambda \in D^0_P(n)  } \mathcal F_{\lambda} \cong V_{D^0_P(n)} \otimes \mathcal H(d)
\]
and
\[
\cB(n)_Q := \bigcap_{j=1}^\ell \text{ker}_{W_{D^0_P(n)}} e_0^{-\alpha_j/\sqrt{n}}.
\]
The claim is that the character of $\cB(n)_Q$ in the instance of $Q=A_{N-1}$ coincides with the Schur-Index of $\mathcal I(N, n)(q, x)$. 
It remains to compute the character. Firstly, 
\[
\ch[\cW(n, \nu)_Q] = q^{\frac{\rho^2}{2n}}\sum_{\lambda \in P^+\cap(Q+\nu)} q^{\frac{n}{2}(\lambda+\rho)^2} \chi_\lambda(z) \frac{1}{\eta(q)^\ell} \sum_{w\in W} \epsilon(w) q^{-(\rho, w(\lambda+\rho))} 
\]
with $\nu \in P$ and 
\[
\cW(n, \nu)_Q := \bigcap_{j=1}^\ell \text{ker}_{V_{\sqrt{n}(Q+\nu)}} e_0^{-\alpha_j/\sqrt{n}}.
\]
Let $m_\lambda(\mu)$ the multiplicity of $\mu$ in the highest-weight representation of highest-weight $\lambda$, so that
\[
\chi_\lambda(z) = \sum_{\mu \in P} m_\lambda(\mu) z^\mu.
\] 
The character of the subspace (actually it is a module for the $U(1)^\ell$-orbifold of $\cW(n)_Q$) 
\[
\cM(n, \mu)_Q := \bigcap_{j=1}^\ell \text{ker}_{\mathcal F_{\sqrt{n}\mu}} e_0^{-\alpha_j/\sqrt{n}}
\]
is the coefficient of $z^\mu$ in the characer of $\cW(n, \nu)_Q$ for $\nu=\mu \mod Q$. 
We thus have
\[
\ch[\cM(n, \mu)_Q]= q^{\frac{\rho^2}{2n}}\sum_{\lambda \in P^+\cap(Q+\mu)} q^{\frac{n}{2}(\lambda+\rho)^2} m_\lambda(\mu) \frac{1}{\eta(q)^\ell} \sum_{w\in W} \epsilon(w) q^{-(\rho, w(\lambda+\rho))}. 
\]
By construction
\[
\cB(n)_Q \cong  \bigoplus_{\mu \in P} \cM(n, \mu)_Q \otimes \mathcal F_{\sqrt{-n}\mu}
\]
and so 
\begin{equation}\nonumber
\begin{split}
\ch[\cB(n)_Q](q,z)&= \sum_{\mu \in  P} \ch[\cM(n, \mu)_Q] \ch[\mathcal F_{\sqrt{-n}\mu}]\\
&= \frac{q^{\frac{\rho^2}{2n}}}{\eta(q)^{2\ell}}\sum_{\mu \in  P} \sum_{\lambda \in P^+\cap(Q+\mu)} q^{\frac{n}{2}(\lambda+\rho)^2} m_\lambda(\mu) \sum_{w\in W} \epsilon(w) q^{-(\rho, w(\lambda+\rho))}  q^{-\frac{n}{2}\mu^2}z^\mu\\
&= \frac{q^{\frac{\rho^2}{2n}(1+n^2)}}{\eta(q)^{2\ell}}\sum_{\lambda \in P^+} q^{\frac{n}{2}(\lambda, \lambda+2\rho)} \sum_{\mu \in  Q+\lambda}   m_\lambda(\mu)q^{-\frac{n}{2}\mu^2} z^\mu\sum_{w\in W} \epsilon(w) q^{-(\rho, w(\lambda+\rho))}.
\end{split}
\end{equation}
The central charge of the $\cB(n)_Q$-algebra is 
\[
c= 2\ell - \frac{\rho^2}{2n}(n-1)^2,
\]
and we check that this precisely matches the proposed central charge  (see \cite{Xie:2012hs}) of the expected chiral algebra for the Argyres Douglas theory in the instance that $Q=A_{N-1}$.
We see that 
\begin{equation}
\ch[\cB(n)_{A_{N-1}}](q,x) = q^{-\frac{c}{24}} \mathcal I(N, n)(q, x)
\end{equation}
or put differently
\begin{equation}
 \mathcal I(N, n)(q, x) = \tr_{\cB(n)_{A_{N-1}}} \left( q^{L_0} \right).
\end{equation}

\section{Another construction and more vertex algebras}\label{sec:morevoas}

There is a second point of view and this one relates to S-duality for \voas{} as explored by Davide Gaiotto and myself \cite{Creutzig:2017uxh}. For this let $\mathfrak g$ be a simply-laced Lie algebra, $Q$ its root lattice and $n$ a positive integer. Let $h^\vee$ be the dual Coxeter number of $\mathfrak g$ and let $k, \ell$ be complex numbers, such that $\psi=k+h^\vee$ and $\psi'=\ell+h^\vee$ and such that
\[
\frac{1}{\psi} + \frac{1}{\psi'} =n.
\]
Then one of our basic proposals is that
\[
A^n[G, \psi] := \bigoplus_{\lambda \in  P^+ \cap\, Q }  \AL_k(\lambda) \otimes \AL_\ell(\lambda)
\]
can be given the structure of a simple \voa{} for generic $\psi$. We can generalize this by including $W$-algebras. For this let $f$ be a nilpotent element of $\mathfrak g$ and denote by $H_{DS, f}$ the corresponding quantum Hamiltonian reduction functor with resulting $W$-algebra $W_k(\g, f) = H_{DS, f}(\AL_k(\g))$. Then we can for example apply this functor to the second factor of $A^n[G, \psi]$ to get
\[
A^n[G, f, \psi] := \bigoplus_{\lambda \in  P^+ \cap\, Q }  \AL_k(\lambda) \otimes H_{DS, f}\left(\AL_\ell(\lambda)\right).
\]
We again expect the existence of the structure of a simple \voa{} on this object, but presently this is only known for $n=1$ and $f$ principal nilpotent by the main Theorem of \cite{Arakawa:2018iyk}.
Let us now take the limit $\psi$ to infinity. Then we expect that we can scale fields in such a way that the limit exists and that $\AL_k(\g)$ becomes in the limit just a large commutative Heisenberg \voa{} $Z$  but the action of the zero-modes survives and integrates to an action of the compact Lie group $G$ of $\mathfrak g$ so that we obtain 
\[
\lim_{\psi\rightarrow \infty} A^n[G, f, \psi] = Z \otimes  \bigoplus_{\lambda \in  P^+ \cap\, Q }  \rho_\lambda \otimes H_{DS, f}\left(\AL_\ell(\lambda)\right)
\]
with $\rho_\lambda$ the irreducible highest-weight representation of highest weight $\lambda$ of $G$.
In the case of $f$ the prinipal nilpotent element the right-hand side coincides as a $G\otimes W_\ell(\mathfrak g, f_{\text{prin}})$-module with $Z\otimes W(p)_Q$. 
This construction is conjectural but we are working on understanding it and for example the cases of $\mathfrak g= \sltwo$ and $n= 1, 2$ are completely understood in \cite{Creutzig:2017uxh, Creutzig:2018ltv}.

Now, I want to employ the picture that has been developped with Kanade and McRae, namely that such \voa{} extensions are possible if and only if one has a braid-reversed equivalence of involved module categories \cite{CKM, CKM2}. Let me explain this in a few words. Assume that we have two \voas{} $V, W$ and sets $V_i, W_i$ of inequivalent simple $V=V_0, W=W_0$ modules such that
\[
A = \bigoplus_i V_i \otimes W_i 
\]
is a \voa{} extension of $V\otimes W$. Then under certain assumptions on the underlying categories $\cC_V, \cC_W$ of $V$ and $W$-modules this implies
that there is a braid-reversed equivalence between $\cC_V$ and $\cC_W$ mapping $V_i$ to the dual of $W_i$. The large $\psi$-limit is a semiclassical limit and there we have to replace one \voa{} by the compact Lie group $G$. Nonetheless we still expect a braid-reversed equivalence to hold \cite{McRae}. 
On the other hand again under certain assumptions on the involved module categories 
\[
A = \bigoplus_i V_i \otimes W_i 
\]
can be given the structure of a simple \voa{} only if there is a braid-reversed equivalence between $\cC_V$ and $\cC_W$ mapping $V_i$ to the dual of $W_i$. In other words, we can now take Rep$(G)$ and replace it by a category $\cC$ equivalent to it and consider the extension
\[
\bigoplus_{\lambda \in  P^+ \cap\, Q }  \cF(\rho_\lambda) \otimes H_{DS, f}\left(\AL_\ell(\lambda)\right)
\]
where $\cF$ is the equivalence from Rep$(G)$ to $\cC$.
So let us construct a category equivalent to Rep$(G)$. Actually we only construct one that is equivalent to the subcategory of Rep$(G)$ whose objects have weights lying in the root lattice and we do this in such a way that we can recover certain subalgebras of $\cB(n)_Q$.
If we forget about the $G$-action and just keep the action of the maximal torus $T$ of $G$, then this forgetful functor embeds Rep$(G)$ into the category of $T$-graded vector spaces. But there is a \voa{} that has tensor categories inside $T$-graded vector spaces, namely the Heisenberg \voa{} of rank the rank of $\mathfrak g$. The identification goes as follows. Let $n$ be a positive integer as before and let $m_\lambda(\mu)$ be the multiplicity of $\mu$ in $\rho_\lambda$, then we define the module 
\[
\cF(\rho_\lambda) := \bigoplus_\mu  \CC^{m_\lambda(\mu)} \otimes_\CC\mathcal F_{\sqrt{-n}\mu}.
\]
The braiding on the $\cF(\rho_\lambda)$ is trivial for weights that lie in the root lattice $Q$ and otherwise only depend on the coset of the weight lattice $P$ in the root lattice $Q$. 
We immediately see that in the case that $f$ is principal nilpotent that 
\[
\bigoplus_{\lambda \in  P^+ \cap\, Q }  \cF(\rho_\lambda) \otimes H_{DS, f}\left(\AL_\ell(\lambda)\right) \subset \cB(n)_Q.
\]
We conjecture that
\[
\cB(n)_{f, Q} := \bigoplus_{\lambda \in  P^+ }  \cF(\rho_\lambda) \otimes H_{DS, f}\left(\AL_\ell(\lambda)\right) 
\]
can be given the structure of a simple \voa{} for any nilpotent element $f$. This can be viewed as a natural generalization of our construction of $\cB(n)_Q$.
Proving this conjecture would amount to proving a generalization of the ideas of Feigin and Tipunin \cite{Feigin:2010xv} to any quantum Hamiltonian reduction. This is difficult. 

Let me summarize this discussion. Let $n$ be a positive integer and $\ell= -h^\vee +\frac{1}{n}$ with $h^\vee$ the dual Coxeter number of the simply-laced Lie algebra $\mathfrak g$ with root lattice $Q$ and weight lattice $P$. Let $f$ be a nilpotent element of $\g$ and let $H_{DS, f}$ be the corresponding quantum Hamiltonian reduction functor. Then conjecturally the following $G \otimes H_{DS, f}(\AL_\ell(g))$-modules have the structure of a simple \voa
\begin{equation}
\begin{split}
\cW(n)_{f, Q} :&= \bigoplus_{\lambda \in  P^+ \cap\, Q }  \rho_\lambda \otimes H_{DS, f}\left(\AL_\ell(\lambda)\right) 
\end{split}
\end{equation}
and the $\mathcal H \otimes H_{DS, f}(\AL_\ell(g))$-modules have the structure of a simple \voa
\begin{equation}
\begin{split}
\cB(n)_{f, Q} :&= \bigoplus_{\lambda \in  P^+ }  \cF(\rho_\lambda) \otimes H_{DS, f}\left(\AL_\ell(\lambda)\right) 
\end{split}
\end{equation}
as well. Here $\mathcal H$ denotes the Heisenberg \voa{} of rank the rank of $\g$. 
We see that by construction the autoorphism group of $\cW(n)_{f, Q}$ contains $G$, while the one of $\cB(n)_{f, Q}$  contains the Weyl group $W$ of $G$ as group of automorphisms as these permute the Fock modules appearing in each $\cF(\rho_\lambda)$. 

Let now $\g=\mathfrak{sl}_n$ and $Y_f$ be the Young tableau corresponding to the nilpotent element $f$. 
Comparing with equation \eqref{eq:01}--\eqref{eq:03} and
for the trivial nilpotent element $0$ one obtains 
\begin{equation}
\begin{split}
\ch[\cB(n)_{0, Q}](z, q) :&= \sum_{\lambda \in  P^+ }   \sum_{\mu \in Q+\lambda} m_{\lambda}(\mu) q^{-\frac{n}{2}\mu^2} x^\mu
 \ch[\AL_\ell(\lambda)](z, q)\\
 &=  q^{-\frac{c}{24}} \sum_{\lambda \in  P^+ }   \tilde f_R^{(n)}(q, x) C^{\Box\Box\cdots\Box}_{\rho_\lambda}(z, q)\\
 &= q^{-\frac{c}{24}}\mathcal I_{A_{N-1, N(n-1)-1}, \Box\Box\cdots\Box}(z, q)
\end{split}
\end{equation}
with $c= (N^2-1)(1-nN) +N-1$ the central charge of $\cB(n)_{0, Q}$. According to the principle of \cite{Beem:2014rza} that the super conformal index associated to the Young tableay $Y_f$ differs from the one of the trivial one $Y_0$ by replacing characters by the corresponding Euler-Poincar\'e characters we immediately get the identification
\begin{equation}
\begin{split}
\ch[\cB(n)_{f, Q}](z, q) :&= \sum_{\lambda \in  P^+ }   \sum_{\mu \in Q+\lambda} m_{\lambda}(\mu) q^{-\frac{n}{2}\mu^2} x^\mu
 \ch[H_{DS, f}\left(\AL_\ell(\lambda)\right)](z, q)\\
 &=  q^{-\frac{c_f}{24}} \sum_{\lambda \in  P^+ }   \tilde f_R^{(n)}(q, x) C^{Y_f}_{\rho_\lambda}(z, q)\\
 &= q^{-\frac{c_f}{24}}\mathcal I_{A_{N-1, N(n-1)-1}, Y_f}(z, q).
\end{split}
\end{equation}
with $c_f$ the central charge of $\cB(n)_{f, Q}$.

\section{Conformal embeddings versus RG flow}\label{sec;emb}

Buican and Nishinaka also discuss RG-flow of Argyres-Douglas theories \cite[Section 4]{BN}. They find that RG-flow interpolates between the Argyres-Douglas theories as
\begin{equation}
(A_{N-1}, A_{N(n-1)-1}) \rightarrow (A_{\nu-1}, A_{\nu(n-1)-1}) \oplus (A_{N-\nu-1}, A_{(N-\nu)(n-1)-1}) \oplus (A_{1}, A_{1}).
\end{equation}
I want to conclude this work with the observation that the chiral algebra of the left-hand side almost embeds conformally in the one of the right-hand side. Here almost means that we have to replace the $\beta\gamma$ \voa, that is the chiral algebra of the $(A_{1}, A_{1})$-theory by a larger \voa.

Let $Q_N=A_{N-1}$ denote the root lattice of $\mathfrak{sl}_N$ and $P_N=A_{N-1}'$ its weight lattice. 
We then have the embeddinng
\[
Q_\nu \oplus Q_{N-\nu} \subset Q_N
\]
and the orthogonal complement is denoted by 
\[
\Phi=\mathbb Z x, \qquad x=(N-\nu)(\epsilon_1 + \dots + \epsilon_\nu) - \nu(\epsilon_{\nu+1} + \dots + \epsilon_N).
\]
We compute that $x^2 = N\nu(N-\nu)$. So that the dual lattice is $\Phi' = \frac{x}{N\nu(N-\nu)}\mathbb Z$ and we have the embedding of dual lattices
\[
P_N \subset P_\nu \oplus P_{N-\nu} \oplus \Phi'.
\]
This induces embeddings of 
\[
\sqrt{n}P_N \oplus \sqrt{-n}P_N \subset  \sqrt{n}P_\nu \oplus \sqrt{-n}P_\nu \oplus \sqrt{n}P_{N-\nu} \oplus \sqrt{-n}P_{N-\nu} \oplus \sqrt{n}\Phi' \oplus \sqrt{-n}\Phi'
\]
and via restriction to the isotropic sublattices defined in \eqref{eq:isotropiclattice}
\[
D^0_{P_N}(n) \subset   D^0_{P_\nu}(n) \oplus D^0_{P_{N-\nu}}(n) \oplus D^0_{\Phi'}(n).
\]
We thus get the following chain of conformal embeddings
\begin{equation}
\begin{split}
\cB(n)_{Q_N} &=   \bigcap_{j=1}^{N-1} \text{ker}_{W_{D^0_{P_N}(n)}} \left(e_0^{-\alpha_j/\sqrt{n}}\right)\\
&\subset   \bigcap_{\substack{ j=1 \\ \j\neq \nu}}^{N-1} \text{ker}_{W_{D^0_{P_N}(n)}} e_0^{-\alpha_j/\sqrt{n}}\\
&\subset   \bigcap_{\substack{ j=1 \\ \j\neq \nu}}^{N-1} \text{ker}_{W_{D^0_{P_\nu}(n)} \otimes W_{D^0_{P_{N-\nu}}(n)} \otimes W_{D^0_{\Phi'}(n)} }\left( e_0^{-\alpha_j/\sqrt{n}}\right)\\
&=  \bigcap_{ j=1}^{\nu-1} \text{ker}_{W_{D^0_{P_\nu}(n)}} \left(e_0^{-\alpha_j/\sqrt{n}}\right)  \otimes \bigcap_{ j=\nu+1}^{N-1} \text{ker}_{W_{D^0_{P_{N-\nu}}(n)}} \left(e_0^{-\alpha_j/\sqrt{n}}\right) \otimes W_{D^0_{\Phi'}(n)}  \\
&= \cB(n)_{Q_\nu} \otimes \cB(n)_{Q_{N-\nu}} \otimes W_{D^0_{\Phi'}(n)}.
\end{split}
\end{equation}
In the last line, we have the chiral algebras of $ (A_{\nu-1}, A_{\nu(n-1)-1})$ and $(A_{N-\nu-1}, A_{(N-\nu)(n-1)-1})$ Argyres-Douglas theories. But the last factor is different than the $\beta\gamma$ \voa. Note however that $\beta\gamma \subset W_{D^0_{\Phi'}(n)}$ by the construction of  \cite{CRW}.

\bibliographystyle{JHEP}

\providecommand{\href}[2]{#2}\begingroup\raggedright\endgroup

\end{document}